\documentclass[aps,prx,twocolumn,showpacs,superscriptaddress,floatfix,nofootinbib]{revtex4-1}
\usepackage{amsfonts}
\usepackage{graphicx}
\usepackage{amsmath}
\usepackage{amssymb}
\usepackage{txfonts}

\usepackage{color}
\usepackage[colorlinks,bookmarks=false,citecolor=blue,linkcolor=blue,urlcolor=blue]{hyperref}

\begin{document}

\title{
Tetramerization in a SU(4)-Heisenberg model on the honeycomb lattice
}

\author{Mikl\'os Lajk\'o}
\affiliation{Institute for Solid State Physics and Optics, Wigner Research
Centre for Physics, Hungarian Academy of Sciences, H-1525 Budapest, P.O.B. 49, Hungary}
\affiliation{Department of Physics, Budapest University of Technology and Economics, 1111 Budapest, Hungary}

\author{Karlo Penc}
\affiliation{Institute for Solid State Physics and Optics, Wigner Research
Centre for Physics, Hungarian Academy of Sciences, H-1525 Budapest, P.O.B. 49, Hungary}
\affiliation{Department of Physics, Budapest University of Technology and Economics, 1111 Budapest, Hungary}
\affiliation{Institute for Solid State Physics, University of Tokyo, Kashiwa 277-8581, Japan}

\date{\today}

\begin{abstract}
The SU(4) Heisenberg model can serve as a low energy model of the Mott insulating state in materials where the spins and orbitals are highly symmetric, or in systems of alkaline-earth atoms on optical lattice. 
Recently, it has been argued that on the honeycomb lattice the model exhibits a unique spin-orbital liquid phase with an algebraic decay of correlations [P. Corboz {\it et al.}, Phys. Rev. X {\bf 2}, 041013 (2012)]. 
Here we study the instability of the algebraic spin-orbital liquid toward spontaneous formation of SU(4) singlet plaquettes (tetramerization). 
Using a variational Monte Carlo approach to evaluate the projected wave-function of fermions with $\pi$-flux state, we find that the algebraic liquid is robust, and that a small finite value of the next nearest exchange is needed to induce tetramerization.
We also studied the phase diagram of a model which interpolates between the nearest neighbor Heisenberg model and a Hamiltonian for which the singlet-plaquette product state is an exact ground state.
\end{abstract}

\maketitle

%

\section{Introduction}

In transition metal oxides the partially filled $d$-orbitals usually undergo a Jahn-Teller distortion which removes the orbital degeneracy. However, in certain cases the orbitals remain fluctuating down to low temperatures at energies comparable to those of spin fluctuations. Possible examples include the triangular system LiNiO$_2$ \cite{kitaoka1998,reynaud2001}, and more recently Ba$_3$CuSb$_2$O$_9$~\cite{nakatsuji2012,quilliam2012} with Cu ions forming a decorated honeycomb lattice. Theoretically, the scenario of spin-orbital liquids was studied in Refs.~\cite{ishihara1997,feiner1997,li1998,khaliullin2000,vernay2004,wang2009,chaloupka2011,nasu2012}.

The minimal model which includes the interaction of spins and orbitals is the Kugel-Khomskii model \cite{kugel1982}. Due to the spatial extension of the orbitals, the Kugel-Khomskii model is usually anisotropic in the orbital part. Under specific circumstances the model acquires its highest symmetry form and can be written as
\begin{equation}
{\cal H}=\sum_{\langle i,j\rangle} 
\left(2 \mathbf{S}_i \cdot \mathbf{S}_j + \frac{1}{2}\right)
\left(2 \mathbf{T}_i \cdot \mathbf{T}_j + \frac{1}{2}\right) \;,
\label{eqn:P2Ham}
\end{equation}
where $\mathbf{S}_i$ are spin-1/2 operators acting on the $\mid\uparrow\rangle$ and $\mid\downarrow\rangle$ states and $\mathbf{T}_j$ are operators acting on the twofold degenerate orbital degrees of freedom (which we denote by $a$ and $b$). For this special case the system is fully SU(4) symmetric. Denoting the local spin-orbital states by colors,  
$\vert {\color{red}\medbullet}  \rangle = \mid \uparrow \! a \rangle$,
$\vert {\color{green}\medbullet} \rangle = \mid \downarrow \! a  \rangle$, 
$\vert {\color{blue}\medbullet} \rangle = \mid \uparrow \! b \rangle$, and
$\vert {\color{yellow}\medbullet} \rangle = \mid \downarrow \! b \rangle$, 
the Hamiltonian (\ref{eqn:P2Ham}) is equivalent to
\begin{equation}
\mathcal H = \sum_{\langle i,j\rangle}  P_{i,j} \;,
\label{eq:HSU4}
\end{equation}
where  $ P_{i,j}$, acting in the color space, interchanges the states on sites $i$ and $j$. This Hamiltonian defines the SU(4) symmetric Heisenberg model. 

The same model also arises in the Mott insulating state of ultra-cold alkaline-earth atoms trapped in an optical lattice. In this case, the nuclear spin of length F becomes the only relevant degree of freedom with $N=2F+1$ states, and their interaction is described by the SU(N) symmetric Heisenberg model\cite{wu2003,honerkamp2004,cazalilla2009,gorshkov2010,taie2012}.

The absence of spin and orbital ordering in Ba$_3$CuSb$_2$O$_9$ \cite{nakatsuji2012,quilliam2012} prompted authors of Ref.~\cite{Corboz12_su4} (including the present authors) to study the symmetric Kugel-Khomskii model on the honeycomb lattice. Combining the results of a variety of numerical and analytical methods, they proposed that an algebraic spin-orbital liquid (or color liquid) state is realized. In this color-liquid the color-color correlation decays algebraically with the distance, and the ground state does not break any space group symmetry, nor the SU(4) symmetry.  In particular, it can be described by a Gutzwiller projected variational wave function based on the $\pi$-flux state of fermions on the honeycomb lattice at 1/4 filling. 
In the $\pi$-flux state a gauge field corresponding to a magnetic flux of half the flux quantum is introduced. The Fermi surface shrinks to a point, as the Fermi-level is at the Dirac point introduced by the $\pi$-flux. 
Naturally, the question arises if the variational state is stable against the opening of a gap at the Fermi level of the unprojected fermions. This can be realized by modulating the hoppings and the on-site energies. Here, our aim is to study the stability of the algebraic color liquid state against such an instability, the tetramerization, using the variational approach.

  The motivation behind choosing tetramerization is not only because it opens a gap in the unprojected fermion spectrum, but also by the natural tendency of the SU(4) colors to form 4-site singlet plaquettes \cite{li1998}. 
  Such plaquettes were observed {\it e.g.} in two-leg SU(4) ladders \cite{vandenBossche2001}, and even a Hamiltonian based on projection operators can be constructed where decoupled SU(4) singlet plaquettes are the exact ground state \cite{chen2005}. Tetramerization also happens in two-dimensional models: on the checkerboard lattice the plaquette-product wave function is an exact ground state of a Hamiltonian which is a sum of projectors \cite{arovas2008}, and the singlet formation persist also in the simple Heisenberg model \cite{Corboz12_simplex}. For the square lattice, the situation is less evident: plaquette formation was proposed in Refs.~\cite{vandenBossche2000,szirmai2011,Hung2011}, but debated in Ref.~\cite{wang2009}. Our current understanding is that a dimerization occurs \cite{corboz11-su4}.
 
The paper is organized as follows: In Sec.~\ref{sec:ProjectionHamiltonian} we construct a Hamiltonian which has singlet-product states as exact ground states. In Sec.~\ref{sec:OrderParameter} we construct an order parameter to detect tetramerization. In Sec.~\ref{sec:VMC} we present results of the variational Monte Carlo calculations. Finally we conclude in Sec.~\ref{sec:Conclusions}.

\section{A model with an exact, tetramerized ground state} 
\label{sec:ProjectionHamiltonian}

\begin{figure}[bth]
  \centering 
  \includegraphics[width=8.5truecm]{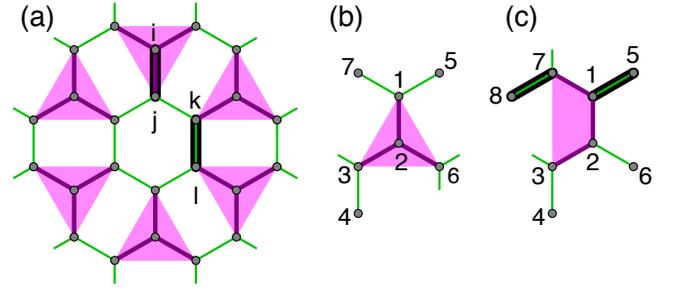}
 \caption{(Color online) (a) The modulation of the bond energies in the tetramerized state on the honeycomb lattice. In the ground state of the $\mathcal{H}_Q$ the spins on the four sites connected by the  purple bonds make a SU(4) singlet (here emphasized by the magenta triangles), these singlet tetramers are decoupled from each other. The two thick black bonds represent the configuration of the $(ij)$ and $(kl)$ bonds of the $Q_{(ij),(kl)}$ projector in the  $\mathcal{H}_Q$ Hamiltonian.  (b) A SU(4) spin singlet that is an eigenstate of the $Q$ projection operators with $0$ eigenvalue. (c) An invalid SU(4) singlet, as the $Q_{(15)(78)}$ cannot be satisfied. 
 \label{fig:tetracov}}
\end{figure}

Using projection operators, we can construct a Hamiltonian for which a singlet-product wave function is an exact ground state, where the four spins of the SU(4) singlets are located on 4-site stars (a central site surrounded by the 3 nearest neighbor sites). There are four such  equivalent coverings, related by translations. For that reason, we consider the 
\begin{equation}
  Q_{(ij),(kl)} = \frac{1}{4} (1+P_{ij})(1+P_{kl})
  \label{eq:Q}
\end{equation}
operator, where $P_{ij}$ is the exchange operator.  As for disjunct $(ij)$ and $(kl)$ the $Q_{(ij),(kl)} =  Q_{(ij),(kl)}^2$ holds, the $Q_{(ij),(kl)}$ is a projection operator with eigenvalues 0 and 1. It gives 0 if the wave function on the $(ij)$ or $(kl)$ bonds is antisymmetric, {\it i.e.} belongs to a singlet. If we choose the $(ij)$ and $(kl)$ bonds to be the nearest neighbor parallel bonds, as shown in Fig.~\ref{fig:tetracov}(a), at least one of the $(ij)$ and $(kl)$ bonds is covered by a SU(4) singlet, so that  
$ Q_{(ij),(kl)} |\Psi_{\text{SP}}\rangle = 0$ for any of the four  $|\Psi_{\text{SP}}\rangle$ tetramer coverings. 
It follows, that the $|\Psi_{\text{SP}}\rangle$ are ground states of the 
\begin{equation}
 \mathcal{H}_Q = \sum_{(ij),(kl)} Q_{(ij),(kl)}
 \label{eq:HQ}
\end{equation} 
Hamiltonian with 0 energy: since it is a sum of positive definite operators, the energy of the ground state must be $\geq 0$. The sum in the $\mathcal{H}_Q$ is over the $3N$ nearest neighbor parallel bond configuration. We note that this construction is slightly different from those mentioned in Refs.~\cite{arovas2008,chen2005,greiter2007}.

We can also easily convince ourselves that no other type of SU(4) singlet coverings satisfies $\mathcal{H}_Q$. Here we follow the site numbering shown in Fig.~\ref{fig:tetracov}: (i) To make $Q_{(12)(34)}$ satisfied, we choose to antisymmetrize the spins on bond $(12)$; (ii) To satisfy $Q_{(15)(23)}$, we antisymmetrize $(23)$; (iii) To make the $Q_{(17)(26)}$ happy, we need to choose to antisymmetrize spins on band $(26)$ [as shown in Fig.~\ref{fig:tetracov}(b)]. Antisymmetrizing $(17)$ would mean that we are not able to satisfy $Q_{(15)(78)}$
[Fig.~\ref{fig:tetracov}(c)], as we cannot antisymmetrize colors on more than four sites. So, the star shaped singlet in  Fig.~\ref{fig:tetracov}(b) is the only possible SU(4) singlet that satisfies all the $Q_{(ij)(kl)}$. The choices we made in step (i) and (ii) in selecting the bonds of $Q$ on which we antisymmetrized the spin lead to the fourfold degeneracy of the ground state.

The following model interpolates between the Heisenberg model and $\mathcal{H}_Q$:
\begin{equation}
 \mathcal{H}_\eta =  \sum_{(ij)} P_{ij}  + \frac{\eta}{4} \sum_{(ij),(kl)} \left( 1 +  P_{ij} P_{kl}\right) \;.
 \label{eq:Ha}
\end{equation} 
For $\eta=1$, the $\mathcal{H}_\eta = \mathcal{H}_Q$, while for $\eta=0$ we recover the nearest neighbor SU(4) Heisenberg model, Eq.~(\ref{eq:HSU4}). To find the transition point between these two cases, we will study this model using the VMC approach in Sec.~\ref{sec:VMC}.

\section{Tetramerization Order parameter}
\label{sec:OrderParameter}

\begin{figure}
  \centering 
  \includegraphics[width=8.5cm]{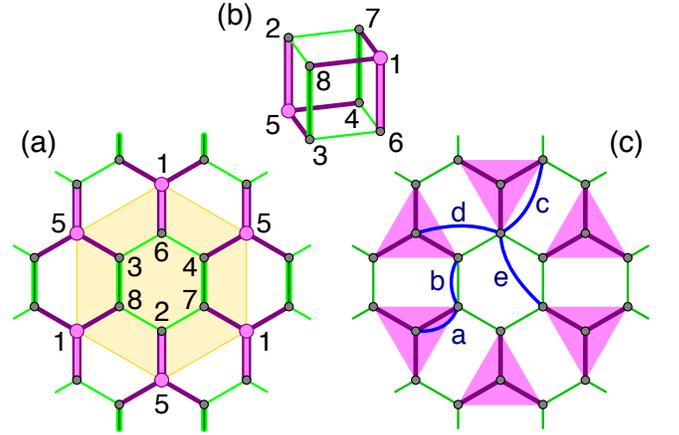}
 \caption{(Color online)
 Graphical representation of the $R_{yz} = P_{1, 6} + P_{2, 5} - P_{3, 8} - P_{4, 7}$ component of the three-dimensional order parameter: (a) on the lattice (the numbers enumerate equivalent sites in the 8-site unit cell), and (b) in the cube.  The cube is topologically equivalent to ordering in the 8-site unit cell. The bonds participating on $R_{xy}$ are shown as double lines.
 We assume that one of out of the four tetramerized ground state is realized, in this case $(R_x,R_y,R_z) \propto (1,1,1)$. (c) The non-equivalent nearest neighbor and next nearest neighbor bond energies in the tetramerized state. In the singlet-product wave functions the $\langle P_a \rangle = \langle P_c \rangle  = -1$ and $\langle P_b \rangle = \langle P_d \rangle =  \langle P_e \rangle  = 1/4$, while in the absence of tetramerization $\langle P_a \rangle = \langle P_b \rangle = \langle P_1 \rangle$ and $\langle P_c \rangle = \langle P_d \rangle =  \langle P_e \rangle = \langle P_2 \rangle$.
\label{fig:ordpar_cube}}
\end{figure} 

To construct the order parameter associated with the tetramerization, we start from the 8 site unit cell of the tetramerized state shown in Fig.~\ref{fig:ordpar_cube}(a). The connectivity of the sites in the unit cell is topologically equivalent to a cube, so the $O_h$ point group of the cube can be used to obtain the order parameter. The expectation values of the $P_{i,j}$ exchange operators on the 12 nearest neighbor bonds in the unit cell transform according to the $\mathsf{A_{1g}}$, $\mathsf{E_{g}}$, $\mathsf{T_{2g}}$, $\mathsf{T_{1u}}$, and $\mathsf{T_{2u}}$ irreducible representations of the $O_h$ group, and can serve to detect order of some kind. The order parameter of the tetramerization is given by
\begin{equation}
\mathbf{R} =
\left(
\begin{array}{c}
P_{1, 6} + P_{2, 5} - P_{3, 8} - P_{4, 7} \\
P_{1, 7} - P_{2, 8} + P_{3, 5} - P_{4, 6} \\
P_{1, 8} - P_{2, 7} - P_{3, 6} + P_{4, 5}
\end{array}
\right) \;,
\end{equation}
and transforms as the three dimensional $\mathsf{T_{2g}}$ irreducible representation. For the singlet product ground state of the $\mathcal{H}_Q$ shown in Fig.~\ref{fig:ordpar_cube}(a) $\mathbf{R} =  (-5/2,-5/2,-5/2)$, while for the remaining three ground states $\mathbf{R} =  (-5/2,5/2,5/2)$, $\mathbf{R} =  (5/2,-5/2,5/2)$, and $\mathbf{R} =  (5/2,5/2,-5/2)$. Denoting the components of the order parameter as  $\mathbf{R} =  (R_{yz},R_{xz},R_{xy})$ [we use the $(yz,xz,xy)$ notation as it also transforms according to the same $\mathsf{T_{2g}}$ irreducible representation in $O_h$], the Landau free energy can be written as 
\begin{equation}
E = c_{R,2} \left(R_{yz}^2 + R_{xz}^2 + R_{xy}^2\right) + 
c_{R,3} R_{yz} R_{xz} R_{xy} + \cdots
\label{eq:LFE}
\end{equation}
The cubic invariant with $c_{R,3}>0$ selects the 4 directions [$R\propto(-1,-1,-1)$, $(-1,1,1)$, $(1,-1,1)$, and $(1,1,-1)$], and makes the phase transition into the tetramerized phase a first order one.

In the variational calculation, where we explicitly break the four-fold degeneracy by hand, it is sufficient to consider the difference of the energies 
on a nearest neighbor  bond connecting two tetramers and on a nearest neighbor bond belonging to a tetramer [see Fig.~\ref{fig:ordpar_cube}(c)]: 
\begin{equation}
r =  \frac{4}{5} \left(\langle P_b \rangle -\langle P_a \rangle \right) \;. 
\label{eq:r}
\end{equation}
For the uniform spin-orbital liquid that does not break the space group of the lattice $r=0$, while for the fully tetramerized state  $ \langle P_{a} \rangle=-1$ and $\langle P_{b} \rangle = 1/4$, so that $r=1$ in this case. 
The two order parameters are related as $\mathbf{R} = 2 (\langle P_a \rangle - \langle P_b \rangle) (1,1,1)=-(5r/2) (1,1,1)$.

\section{Variational Monte Carlo results}
\label{sec:VMC}

The variational approach is based on the fermionic representation of the SU(N) algebra. The exchange operator $P_{i,j}$ can be written as
\begin{equation}
  P_{i,j} = -\sum_{\alpha,\beta=1}^N f^\dagger_\alpha(i) f^\dagger_\beta(j)
  f^{\phantom{\dagger}}_\beta(i) f^{\phantom{\dagger}}_\alpha(j) \;,  
\end{equation}
where $f^\dagger_\alpha(i)$ creates a fermion of color $\alpha$ on site $i$, and $f^{\phantom{\dagger}}_\alpha(i)$ annihilates it. To treat the four fermion term, it is customary to introduce a bond mean-field decoupling of the $P_{i,j}$ \cite{affleck1988-sun}, leading to a free-fermion Hamiltonian, where the hopping amplitudes are determined self-consistently. The mean-field procedure often results in solutions where the product of the hoppings around a plaquette is a complex number $\propto e^{i \Phi}$, as if the fermions were picking up a phase due to a magnetic field with flux $\Phi$ threading through the plaquette (in convenient units). Such a finite flux can significantly change the band structure. Chiral solutions (with $\Phi\neq0, \pi$) were found for the SU(N) model with $N\geq5$ on the square lattice \cite{hermele2009} and for the SU(6) model on the honeycomb lattice \cite{szirmaiG2011}. While the mean-field solutions give a very useful insight into the possible nature of the ground state, it describes free fermions where the on-site occupancy is fixed on the average only. 

 In a complementary approach one does not search for mean-field solutions, but takes the ground state wave function of some free-fermion Hamiltonian, and by applying a Gutzwiller projection one ensures that the occupation of each site is exactly one. The projection is done numerically on a finite size cluster, using Monte-Carlo importance sampling, where one can efficiently sample the wave function and calculate the energy and correlations \cite{yokoyama1987,gros1989}. Regarding two-dimensional SU(N) Heisenberg models, the method has been applied to the SU(4) model on the square \cite{wang2009} and honeycomb lattice \cite{Corboz12_su4}, and to the SU(3) model on the triangular lattice \cite{Bieri2012} and honeycomb lattice \cite{corboz2013_su3_hc}. 
 
 \begin{figure}[t!]
  \centering 
  \includegraphics[width=7.5truecm]{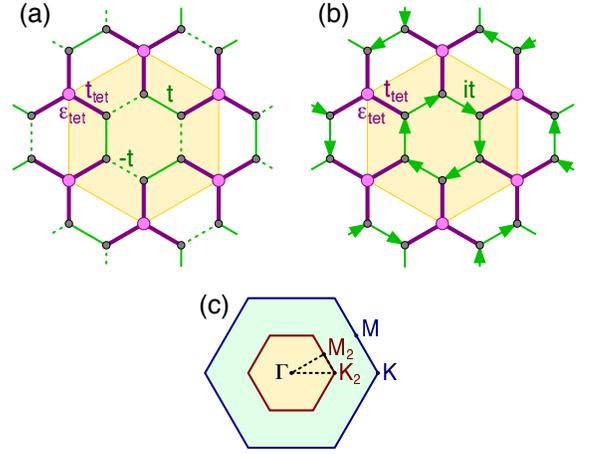}
 \caption{(Color online)  (a) The modulation of the hopping amplitudes and of the on-site energies in the $\pi$-flux hopping Hamiltonian using real hopping amplitudes. The sign of the hopping amplitudes on the solid and dashed green lines are opposite, so that the product of the hoppings over the six bonds of any hexagon is negative. The 8-site hexagonal unit cell is shown in orange. (b) The $\pi$-flux Hamiltonian using real $t_{\text{tet}}$ and imaginary $it$ amplitudes. Since $t_{j,k}=t^*_{k,j}$, the arrows $j\rightarrow k$ represent $t_{j,k}=it$. (c) The Brillouin zone of the honeycomb lattice (blue, with high symmetry points $K$ and $M$) and of the 8-site unit cell (red, with high symmetry points $K_2$ and $M_2$) 
 \label{fig:ttetetet}}
\end{figure}

\begin{figure*}
  \centering 
  \includegraphics[width=15cm]{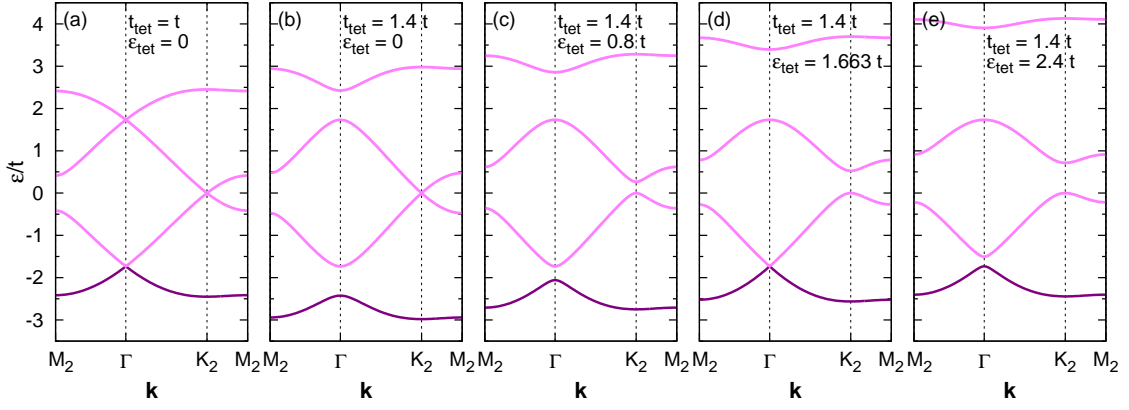}
 \caption{(Color online) The two-fold degenerate bands of the free fermions of the tetramerized $\pi$-flux state for a few selected values of the $\varepsilon_{\text{tet}}$ and $t_{\text{tet}}$ along a path connecting high-symmetry points in the reduced Brillouin zone [see Fig.~\ref{fig:ttetetet}(c)]. $\Gamma$ is the center of the Brillouin-zone (${\bf k}=0$), $M_2$ is the ${\bf k}= (\pi/3, 0)$ (and the symmetry related points), and $K_2$ is the ${\bf k} = (0, 2\pi/{3\sqrt{3}})$. Occupied  bands are colored dark purple, while empty bands are colored light purple. 
 \label{fig:bands}}
\end{figure*}

For the SU(4) model on the honeycomb lattice, the energy of the Gutzwiller projected $\pi$-flux state is much lower than the energy of the $0$-flux state  \cite{Corboz12_su4}.  So to study tetramerization within the variational approach, we modify the free fermion Hamiltonian by strengthening the hoppings $t_{\text{tet}}$ on the bonds belonging to the tetramers and by introducing a negative onsite energy $\varepsilon_{\text{tet}}$ for the sites at the center of the tetramers, while maintaining the $\pi$-flux state. The Hamiltonian is then given by 
\begin{equation}
 \mathcal{H}_{f}  =  \sum_{\alpha=1}^4 \left[
 \sum_{i}  \varepsilon_i
f_{i,\alpha}^{\dagger} f_{i,\alpha}^{\phantom{\dagger}}
 -   
\sum_{\langle i,j \rangle}     
\left(  t_{i,j} 
  f_{j,\alpha}^{\dagger} f_{i,\alpha}^{\phantom{\dagger}} + t_{j,i}  f_{i,\alpha}^{\dagger} f_{j,\alpha}^{\phantom{\dagger}}  \right)
\right] \;,
\label{eq:hopham}
\end{equation} 
where the $t_{i,j}=t^*_{j,i}$ hoppings and $\varepsilon_i$ on-site energies are chosen as shown in Fig.~\ref{fig:ttetetet}(a). The sign of the $t_{i,j}$ is such that the product of the hoppings around the hexagons is negative: the product is either $-t^6$ or $-t^2 t_{\text{tet}}^4$, depending on the hexagon. This realizes the $\pi$-flux state, the variational state with the lowest energy for the nearest neighbor exchange. The uniform solution corresponds to the $t=t_{\text{tet}}$ and $\varepsilon_{\text{tet}}=0$


\subsection{Free fermions in the $k$-space}

Before discussing the results of the Gutzwiller projection, it is instructive to look at the properties of free fermions with $\pi$-flux per plaquette. Fourier transforming the fermionic operators in a conveniently chosen gauge, we can write the Hamiltonian (\ref{eq:hopham}) in the $k$-space as an $8\times 8$ matrix of the form (following the site ordering shown in Fig.~\ref{fig:ordpar_cube})
\begin{equation}
\mathcal{H}_f({\bf k}) = \left(
\begin{array}{cc}
\mathcal{E} & -\mathcal{F}  \\
-\mathcal{F}^\dagger & \mathcal{E} 
\end{array}
\right)\;,
\label{eq:Hf}
\end{equation}
where the only nonzero element of $\mathcal{E}$ matrix is $\mathcal{E}_{1,1}=\varepsilon_{\text{tet}}$ 
and 
\begin{equation}
\mathcal{F} = \left(
\begin{array}{cccc}
 0 & e^{2 i k_y} t_{\text{tet}} & e^{i \sqrt{3} k_x-i k_y} t_{\text{tet}} & e^{-i \sqrt{3} k_x-i k_y} t_{\text{tet}} \\
 e^{2 i k_y} t_{\text{tet}} & 0 & t & -t \\
 e^{i \sqrt{3} k_x-i k_y} t_{\text{tet}} & -t & 0 & t \\
 e^{-i \sqrt{3} k_x-i k_y} t_{\text{tet}} & t & -t & 0 
\end{array}
\right)
\end{equation}
describes the hopping between the nearest neighbor sites. In the absence of the on-site potential $\varepsilon_{\text{tet}}$, the $\mathcal{E}=0$ and by squaring the eigenvalue problem we get  $\mathcal{F}^\dagger \mathcal{F} |\psi\rangle = \varepsilon^2 |\psi\rangle$, {\it i.e.} the bands are particle-hole symmetric.

Solving for the eigenvalues of Eq.~(\ref{eq:Hf}), we find four two-fold degenerate bands in the Brillouin zone of the 8-site unit cell, shown in Fig.~\ref{fig:bands}. The two-fold degeneracy is the direct consequence of the $\pi$-flux. In the limiting case of $t_{\text{tet}}=0$, when we end up with a hexagon and two isolated sites in the unit cell, the degeneracy is present as the $t$ hoppings with alternating signs on the hexagon make the energy levels double degenerate. In the other limit, when $t=0$, the situation is simpler, as there are two decoupled tetramers in the unit cell, trivially giving a two-fold degeneracy. To see the degeneracy for arbitrary values of hopping parameters, we start from a one-fermion wave function $|\psi_{1A}\rangle$, where the amplitude is 1 at the center of one of the tetramers. We can then generate a four-dimensional Hilbert-space by successively acting with the $\mathcal{H}_f({\bf k})$ to the $|\psi_{1A}\rangle$ state:
\begin{eqnarray}
|\psi_{1A}\rangle &=&\left(0, 0, 0, 0, 1, 0, 0, 0\right) \;,
\\
|\psi_{2A}\rangle &=&\frac{1}{\sqrt{3}} \left(0, -1,-1, -1, 0, 0, 0, 0\right)\;,
\\
|\psi_{3A}\rangle &=&\frac{1}{3 \sqrt{1- |\gamma_{\bf k}|^2}}
\left(0,
\gamma_{\bf k}^*-e^{2 i k_y},
\gamma_{\bf k}^*-e^{i \sqrt{3} k_x-i k_y},
\right.\nonumber\\&&\left.
\gamma_{\bf k}^*-e^{-i k_y-i \sqrt{3} k_x},
0,0,0,0\right)\;,
\\
|\psi_{4A}\rangle &=&
\frac{i}{\sqrt{27} \sqrt{1- |\gamma_{\bf k}|^2}}
\left(0,0,0,0,0,
e^{-i \sqrt{3} k_x -i k_y}-e^{i \sqrt{3} k_x-i k_y},
\right.\nonumber\\&&\left.
e^{2 i k_y}-e^{-i \sqrt{3} k_x-i k_y},
e^{i \sqrt{3} k_x-i k_y}-e^{2 i k_y}
   \right)\;,
\end{eqnarray}
where
\begin{equation}
\gamma_{\bf k}=\frac{1}{3} \left(
    e^{-i \sqrt{3} k_x + i k_y}+e^{i \sqrt{3} k_x + i k_y}+e^{-2 i k_y}
  \right) \;.
\end{equation}
Further acting with $\mathcal{H}_f({\bf k})$ on $|\psi_{4A}\rangle $  results in  a linear combination of $|\psi_{1A}\rangle$, $|\psi_{2A}\rangle$, $|\psi_{3A}\rangle$, and $|\psi_{4A}\rangle$, so that the Hilbert space closes.
In this Hilbert space the hopping Hamiltonian reduces to
\begin{equation}
\mathcal{H}_f' = 
\left(
\begin{array}{cccc}
\varepsilon_{\text{tet}} & \sqrt{3} \gamma_{\bf k}^* t_{\text{tet}} & \sqrt{3 \!-\! 3|\gamma_{\bf k}|^2} t_{\text{tet}}  & 0 \\
 \sqrt{3} \gamma_{\bf k}^{\phantom{*}} t_{\text{tet}} & 0 & 0 & 0 \\
 \sqrt{3\!-\!3|\gamma_{\bf k}|^2} t_{\text{tet}} & 0 & 0 & -i \sqrt{3} t \\
 0  & 0 & i \sqrt{3} t & 0
\end{array}
\right)\;.
\label{eq:Hfred}
\end{equation}
The dispersion $\varepsilon_{\bf k}$  of the fermions is given by the characteristic polynomial of $\mathcal{H}_f'$ 
\begin{equation}
0 =
\varepsilon_{\bf k}^4
-\varepsilon_{\text{tet}} \varepsilon_{\bf k}^3 
-3 \varepsilon_{\bf k}^2 \left(t^2+t_{\text{tet}}^2\right)
+ 3 \varepsilon_{\text{tet}} \varepsilon_{\bf k}  t^2
+ 9 |\gamma_{\bf k}|^2 t^2 t_{\text{tet}}^2 \;.
\label{eq:bande}
\end{equation}
Similarly, we can start with $|\psi_{1B}\rangle$, where the amplitude is 1 at site 1. This leads to a four-dimensional Hilbert space that is orthogonal to the $\{|\psi_{1A}\rangle,|\psi_{2A}\rangle,|\psi_{3A}\rangle,|\psi_{4A}\rangle\}$, with elements $|\psi_{jB}\rangle = \mathcal{I} |\psi_{jA}\rangle$, where $\mathcal{I}$ is the inversion operator with center of inversion being the center of the hexagon with the alternating $t$ and $-t$ hopping amplitudes. We end up with a  reduced hopping matrix that is the complex conjugate of $ \mathcal{H}_f'$, with a spectrum given also by Eq.~(\ref{eq:bande}). So we have successfully block diagonalized  $\mathcal{H}_f$ and demonstrated that the spectrum is two-fold degenerate. In fact, using a local gauge transformation $f_{j,\alpha}^\dagger \rightarrow i f_{j,\alpha}^\dagger$ on sites $j=1$,6,7, and 8 of the unit cell transforms the phases of the hoppings in such a way that the inversion symmetry is explicitly manifested, as shown in Fig.~\ref{fig:ttetetet}(b), with a point group symmetry $\mathcal{C}_6$ (the configuration with real hoppings, shown in Fig.~\ref{fig:ttetetet}(a), has the point group symmetry $\mathcal{D}_3$).

In Fig.~\ref{fig:bands} we show the band structure for different choices of the $t_{\text{tet}}$ and $\varepsilon_{\text{tet}}$. The bands of the uniform $\pi$-flux state of the algebraic color liquid are shown in Fig.~\ref{fig:bands}(a). There are three two-fold degenerate Dirac-points in the band structure, two at $\Gamma$ point with $\varepsilon=\pm\sqrt{3}t$ and one at the $K_2$ point with $\varepsilon=0$  - the latter one is counterpart of the Dirac-point of the hopping hamiltonian with uniform hoppings, and remains a Dirac point as long as $\varepsilon_{\text{tet}}=0$, as seen in Fig.~\ref{fig:bands}(b). For the SU(4) problem, however, the lower Dirac-point at ${\bf k}=0$ is more important: it makes the Fermi surface of the quarter filled Fermi sea point-like and is responsible for the algebraic decay of the color-color correlation functions. Tetramerization removes the Dirac point and a gap opens between the lowest band and the band above it at the $\Gamma$ point. However, fine tuning the parameters, we can recover the Dirac point:
for ${\bf k}=0$ the Eq.~(\ref{eq:bande}) factorizes into
$(\varepsilon_{\bf k}^2 - 3 t^2)(\varepsilon_{\bf k}^2 -\varepsilon_{\text{tet}} \varepsilon_{\bf k}  - 3 t_{\text{tet}}^2)=0$. As we change the $\varepsilon_{\text{tet}}$ and $t_{\text{tet}}$, the gap closes at $\varepsilon = -\sqrt{3} t$ when  
\begin{equation}
\varepsilon_{\text{tet}} = \frac{\sqrt{3}}{t}\left(t_{\text{tet}}^2-t^2\right)\;.
\label{eq:evst_dirac}
\end{equation}
A typical spectrum with a Dirac-point for such a particular value of $\varepsilon_{\text{tet}}$ is shown in Fig.~\ref{fig:bands}(d). We note that in a finite size cluster a level crossing happens in the free-fermion spectrum at the $\Gamma$ point.

\subsection{Gutzwiller projection}
 
In the following, we calculate the expectation value of the exchange Hamiltonian in the Gutzwiller projected wave function using Monte-Carlo importance sampling as we change the values of $t_{\text{tet}}$ and $\varepsilon_{\text{tet}}$ in the $\pi$-flux phase. A short description of the method we use is given in Ref.~\cite{Corboz12_su4}. 

Here we considered a family of finite size clusters with $N = 24 L^2$ sites defined by the lattice vectors  ${\bf g}_1 = (3\sqrt{3},3)L$ and ${\bf g}_2 = (0,6)L$ where $L$ is an integer and the distance between the nearest neighbor sites is unity.  We performed calculations on  clusters with $N=24$, 96, 216, and 384 sites ($L = 1$,2,3, and 4). All of these clusters have the full $\mathcal{D}_6$  symmetry of the honeycomb lattice and are compatible with the tetramerization pattern. In our Monte Carlo algorithm an elementary update is the exchange of two randomly chosen fermions at arbitrary sites with different colors. The sampling distances were chosen to be around five times the autocorrelation length, so that each measurement is practically independent. In each run the error of the Monte Carlo results is of the order of $10^{-4}$, thus much smaller than the symbol sizes on the plots. The values of the sampling distances and the  number of measurements for the different custers can be found in Table \ref{table:corrtimes}.

\begin{table}[htdp]
\begin{center}
\begin{tabular}{r|r|r|r|r|r}
N  & $\tau_{\text{a.c.}}$ & $\Delta n_{\text{meas.}}$ & $\Delta n_{\text{meas.}}/\tau_{\text{a.c.}}$ & $n_{\text{meas.}}$ \\
\hline
24 & 22 & 1000 &  45.5 & $10^7$\\
96 & 260 & 1000 & 3.8 &$5 \cdot10^6$\\
216 & 1080 & 5000 & 4.6& $2\cdot10^6$\\
384 & 2920 & 20000 & 6.8&$2\cdot10^6$
\end{tabular}
\end{center}
\caption{Estimated autocorrelation times $\tau_{\text{a.c.}}$ for the nearest neighbor bond energies, the number of elementary step between the measurements ($\Delta n_{\text{meas.}}$), the ratio between $\Delta n_{\text{meas.}}$ and $\tau_{\text{a.c.}}$, and the total number of measurements $n_{\text{meas.}}$ in our Monte-Carlo runs for clusters with $N$ sites (the number of elementary updates is equal to $n_{\text{meas.}}\times \Delta n_{\text{meas.}}$).}
\label{table:corrtimes}
\end{table}


\subsubsection{Stability of the algebraic color liquid}

\begin{figure}
  \centering 
  \includegraphics[width=8.cm]{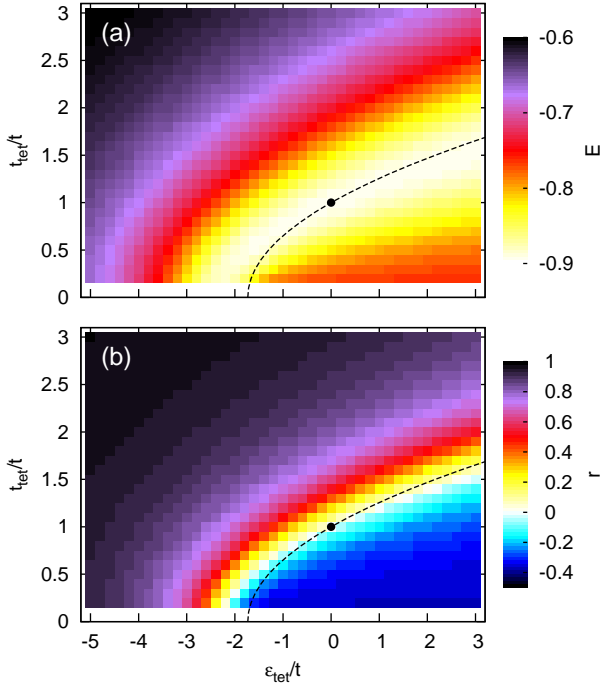}
 \caption{(Color online) (a) The energy of the SU(4) nearest neighbor Heisenberg model and (b) the order parameter $r$ in the Gutzwiller projected variational wave function as a function of the $t_{\text{tet}}$ and $\varepsilon_{\text{tet}}$ parameters of the hopping Hamiltonian, calculated on a 96 site cluster. The $t_{\text{tet}}$ and $\varepsilon_{\text{tet}}$ were varied in steps of $0.1t$ and $0.2t$, respectively.
Along the dashed black line the free fermion Fermi surface becomes a point, following the condition given by Eq.~(\ref{eq:evst_dirac}). The energy minima closely follows the line, and similarly the order parameter vanishes in the vicinity of the line. The non-tetramerized $t_{\text{tet}}=t$, $\varepsilon_{\text{tet}}=0$ case is denoted by a black point.
 \label{fig:quadfig}
 }
\end{figure}

\begin{figure}
  \centering 
  \includegraphics[width=8.5cm]{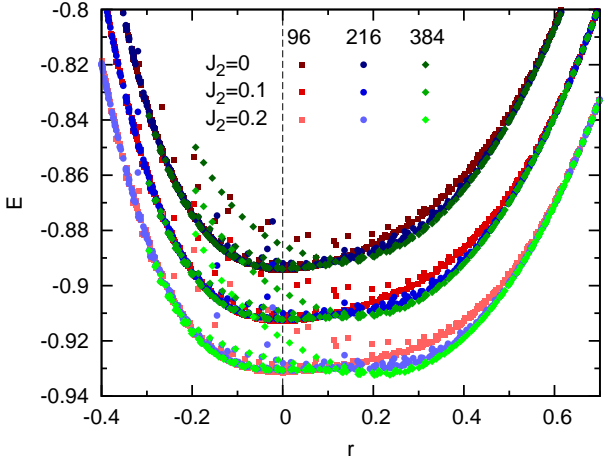}
 \caption{(Color online)
 Energy per site, Eq.~(\ref{eq:Ennn}), for $J_2=0$, 0.1, and 0.2 as a function of the tetramerization order parameter $r$, Eq.~(\ref{eq:r}), in the 96, 216, and 384 site cluster for various $t_{\text{tet}}$ and  $\varepsilon_{\text{tet}}$ values. We set $J_1=1$. The energies are shifted by $-0.2 J_2$ so that the points belonging to different $J_2$ values are easily discerned. Statistical errors are smaller than the symbol sizes. 
  \label{fig:E_vs_r_J2}
 }
\end{figure}

\begin{figure}
  \centering 
  \includegraphics[width=8.cm]{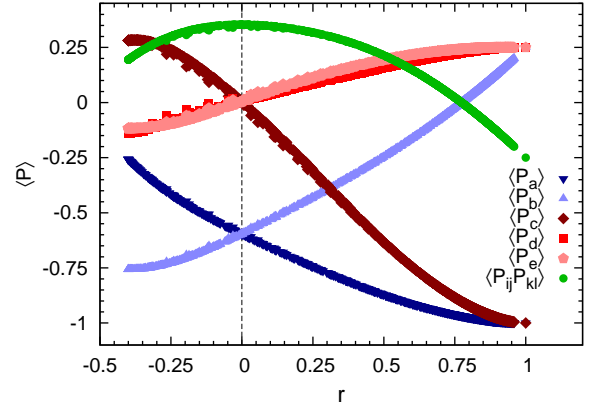}
 \caption{(Color online)
   The expectation values of the exchange operators on the non-equivalent first and second neighbor bonds shown in Fig.~\ref{fig:ordpar_cube}(c), and of the four site $P_{ij}P_{kl}$ operator as a function of the tetramerization parameter $r$, Eq.~(\ref{eq:r}), in the 96 site cluster for various $t_{\text{tet}}$ and  $\varepsilon_{\text{tet}}$ values. Statistical errors are smaller than the symbol sizes. 
  The singlet product wave function is the maximally tetramerized state with $r=1$ and $\langle P_a \rangle = \langle P_c \rangle  = -1$, $\langle P_b \rangle = \langle P_d \rangle =  \langle P_e \rangle  = 1/4$, and   $\langle P_{ij}P_{kl} \rangle = 1/4$.
 \label{fig:E_vs_r0}
 }
\end{figure}

First, we consider the stability of the algebraic color liquid against tetramerization. For that reason we vary the $t_{\text{tet}}$ and $\varepsilon_{\text{tet}}$ in a systematic way and measure the expectation value of the energy and the tetramerization order parameter, [Eq.~(\ref{eq:r})]. The energy per site is given by
\begin{equation}
 E = \frac{3}{2} \langle P_1 \rangle \;.
 \label{eq:Enn}
\end{equation}
By $\langle P_1 \rangle$ we denote the average value of nearest neighbor bond exchange,
\begin{equation}
 \langle P_1 \rangle =  \frac{1}{2} 
 \left(\langle P_a \rangle + \langle P_b \rangle \right) \;,
 \label{eq:P1def}
\end{equation}
where $\langle P_a \rangle$ and $\langle P_b \rangle$ are the expectation value of the exchange on the intra- and inter-tetramer bonds, respectively [Fig.~\ref{fig:ordpar_cube}(c)].
Results for the energy as a function of the $t_{\text{tet}}$ and $\varepsilon_{\text{tet}}$ are shown in Fig.~\ref{fig:quadfig}(a), together with the tetramerization parameter, Fig.~\ref{fig:quadfig}(b). 
It appears that the energy is minimal and the tetramerization order parameter disappears in a narrow valley roughly following the condition (\ref{eq:evst_dirac}) required for the Dirac points to form at the Fermi surface. The actual minimum is realized for the $t_{\text{tet}}=t$ and  $\varepsilon_{\text{tet}}=0$. This is explicitly seen in Fig.~\ref{fig:E_vs_r_J2}, where the energy and order parameter values of Fig.~\ref{fig:quadfig} are replotted with $t_{\text{tet}}$ and  $\varepsilon_{\text{tet}}$ being implicit variables: The $\pi$-flux state is both locally and globally stable against tetramerization for the nearest neighbor exchange on the Heisenberg lattice for the shown cluster sizes. Quite surprisingly, the tuning of the parameters $t_{\text{tet}}$ and $\varepsilon_{\text{tet}}$ in the free fermion Hamiltonian may have similar effects after the Gutzwiller projection. Namely, even for quite different values of $t_{\text{tet}}$ and $\varepsilon_{\text{tet}}$ the $E$ vs.\ $r$ plot essentially lay on a single curve. Deviation from this behavior is seen for small $t_{\text{tet}}$ and not too large values of $|r|$ -- these points are scattered above the principal energy curve on the plot (for larger values of $|r|$ the points for these hoppings merge with other points). This behavior is also seen in Fig.~\ref{fig:E_vs_r0} for the exchanges on non-equivalent bonds. We also checked the nearest neighbor color-color correlations in the unprojected Fermi-sea states, but found no sign of similar behavior, so the collapse of different points onto a single curve is the results of the Gutzwiller projection. 

\begin{figure}
  \centering 
  \includegraphics[width=8.cm]{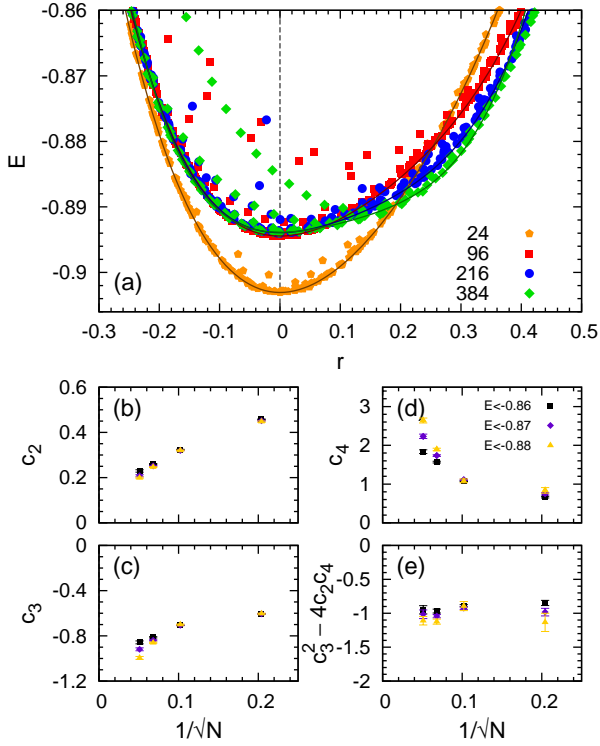}
 \caption{(Color online)
 (a) The energy $E=3 \langle P_1 \rangle /2$ as a function of the tetramerization parameter for 24, 96, 216, and 384 site clusters, together with a fit based on Eq.~(\ref{eq:efit}) using the points that are below the energy threshold -0.87. In (b)--(d) we plot the fitting parameters for the energy thresholds -0.86, -0.87, and -0.88, as a function of $1/\sqrt{N}$, where $N$ is the size of the system. (e) The finite size dependence of the discriminant.
 \label{fig:ctuning}
 }
\end{figure}

Comparing the energy for the different cluster sizes in Fig.~\ref{fig:E_vs_r_J2}, we find that the finite size corrections depend on the value of the tetramerization: while they are negligible for $r=0$, they are significant when $0.1 \lesssim r \lesssim 0.5$, and the energy of the tetramerized state decreases. To check whether the $r=0$ remains the solution in the thermodynamic limit, in Fig.~\ref{fig:ctuning}(a) we plot the energy against the tetramerization order parameter for different cluster sizes (up to 384 sites) in a narrower energy window. To capture the shape and the asymmetry of the energy curve, we fitted the energy by the 
 \begin{equation}
   E_{\text{fit}} = E_0 + c_2 r^2 + c_3 r^3 + c_4 r^4
   \label{eq:efit}
 \end{equation} 
trial form, relying on the  expression of the free energy, Eq.~(\ref{eq:LFE}). The results of the fit for different cluster sizes are shown in Fig.~\ref{fig:ctuning}(b)-(d), together with the size dependence of the fitting parameters $c_2$, $c_3$, and $c_4$. For each system size we made three fits, using points selected by the $E<-0.86$, $E<-0.87$, and $E<-0.88$ criteria. The $c_2$ coefficient has a nice scaling, and in the thermodynamic limit it goes to  $c_2\approx 0.12$, giving evidence for the local stability. For the $c_3$ and $c_4$ the finite size effects are more involved. In fact, since we are primarily interested if the $E_0$ remains the lowest energy state, we followed a different strategy: to confirm that $E_0$ is the lowest energy we examine the sign of $E-E_0= r^2(c_2+c_3r+c_4r^2)$. If it is positive for all $r\neq0$, i.e.\ the discriminant of the $c_2+c_3r+c_4r^2$ is negative, then $r=0$ is clearly the global minimum of Eq.~(\ref{eq:efit}) with energy $E_0$. The phase transition occurs when the discriminant becomes $0$.  We plotted the discriminant as a  function of the inverse size in  Fig.~\ref{fig:ctuning}(e). It appears that the discriminant remains finite and negative even in the $N\rightarrow \infty$ thermodynamic limit, noting again that the finite size correction are difficult to assess. In other words, the variational calculation seems to confirm stability of the algebraic liquid against tetramerization.

\begin{figure}
  \centering 
  \includegraphics[width=8.cm]{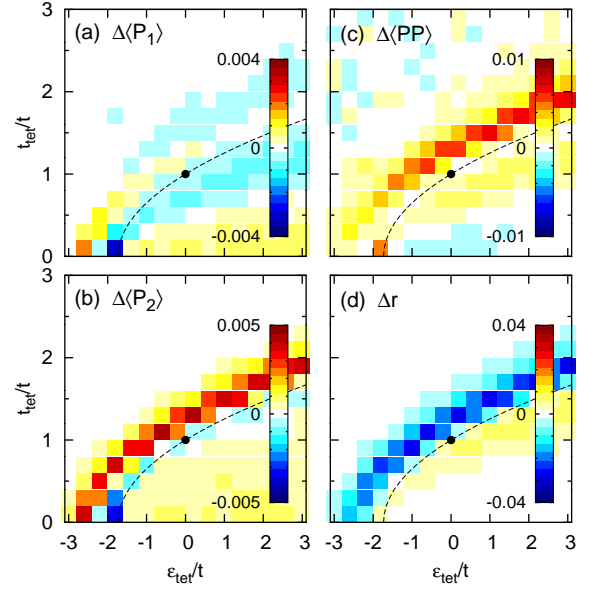}
 \caption{(Color online)
To illustrate the finite size effects, we plot the difference between the 96 cluster and 216 cluster results for the expectation values of the 
(a) nearest neighbor bond exchange average $\Delta \langle P_1 \rangle = \langle P_1 \rangle_{N=96}-\langle P_1 \rangle_{N=216}$, 
(b) second neighbor  bond exchange average $\Delta \langle P_2 \rangle$,
(c) four-site exchange  $\Delta\langle PP \rangle$, and (d) the order parameter.
\label{fig:D_96_216} 
 }
\end{figure}

In Fig.~\ref{fig:D_96_216} we plot the difference between the 96 and the 216 site cluster results of the $\langle P_1 \rangle$, $\langle P_2 \rangle$, $\langle P_{ij}P_{kl}  \rangle$, and $r$  as a function of the hopping parameters $t_{\text{tet}}$ and $\varepsilon_{\text{tet}}$. The $\langle P_1 \rangle$ appears to be the least dependent on the cluster size. For all the other quantities the finite size effects are large for the moderately strong tetramerized case, in a region that is running parallel the dashed line where the band gap closes and a Fermi point appears [Eq.~(\ref{eq:evst_dirac})].

\subsubsection{Phase transition between the algebraic color liquid and the tetramerized state}

\begin{figure}
  \centering 
  \includegraphics[width=8.5cm]{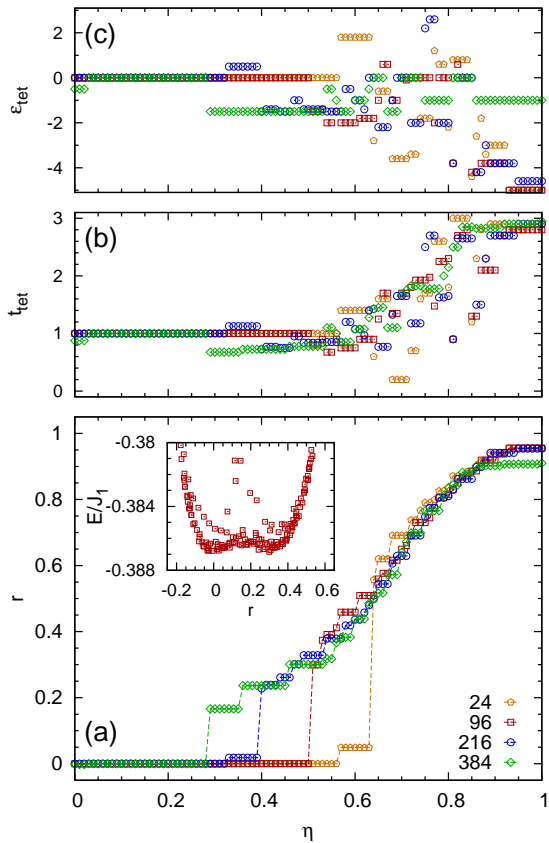}
 \caption{\label{fig:alpha_r}
 (a) The tetramerization order parameter $r$ as a function of $\eta$ in Eq.~(\ref{eq:Ha}) for 24, 96, 216, and 384 site clusters.
  A first-order phase transition occurs between the algebraic color liquid ($r=0$) and the tetramerized phase ($r>0$), the $\eta$-value of the phase transition decreases with increasing cluster size. 
  Since we considered wave functions with finite values of $t_{\text{tet}}$ and $\varepsilon_{\text{tet}}$ in the Monte Carlo sampling, the $r=1$ ground state for $\eta=1$ is not shown in the plot.
   In the inset we show the expectation value as a function of the order parameter close to the phase transition, at $\eta=0.5$ for the 96-site cluster. Each points corresponds to a different value of the free-fermion parameters $t_{\text{tet}}$ and  $\varepsilon_{\text{tet}}$, as calculated by the Variational Monte Carlo on the 96 site cluster. In (b) and (c) we show the values of $t_{\text{tet}}$ and $\varepsilon_{\text{tet}}$  giving the minimal energy solution in the parameter set we considered.}
\end{figure}

For $t=0$ the free-fermion Hamiltonian describes decoupled tetramers with localized fermions. The lowest energy level of a decoupled star is singly degenerate, and Gutzwiller projecting the four fermion ground state gives an SU(4) singlet. Not surprisingly, the Gutzwiller projected wave function becomes the tetramer-factorized wave function in this case, the exact ground state of the of the Hamiltonian (\ref{eq:HQ}). So changing the $\eta$ in the Hamiltonian (\ref{eq:Ha}), we may expect that a phase transition occurs at a finite value of the $\eta$ parameter between the uniform color liquid and the tetramerized state and that the transition can be captured by the variational wave function.

To explore this possibility, we have looked up the minimal energy among the calculated $t_{\text{tet}}$ and $\varepsilon_{\text{tet}}$ values for a given $\eta$, and plotted the corresponding value of order parameter $r$. The energy per site is given by
\begin{equation}
 E = \frac{3}{2} \langle P_1 \rangle +
  \frac{3}{4} \eta \left( 1 +\langle P_{ij}P_{kl} \rangle \right) \;,
\end{equation}
and the result is shown in Fig.~\ref{fig:alpha_r}. For each cluster, the algebraic color liquid is stable up to a value of $\eta$, where a first order transition happens into the tetramerized state -- this agrees with the existence of the cubic invariant of the order parameter in Eq.~(\ref{eq:LFE}).The first order nature of the transition is further supported by the energy versus order parameter plot, shown in the inset of Fig.~\ref{fig:alpha_r} for $\eta=0.5$ and $N=96$ : we can clearly recognize the two minima of the energy, one corresponding to the algebraic color liquid ($r=0$), the other at $r\approx 0.3$ to the tetramerized state. The point of transition decreases with the system size.
 We did not find a reliable finite size extrapolation, so the value of the phase transition in the thermodynamic limit remains uncertain. 
  
\subsubsection{Tetramerization due to next nearest exchange}

\begin{figure}
  \centering 
  \includegraphics[width=8.5cm]{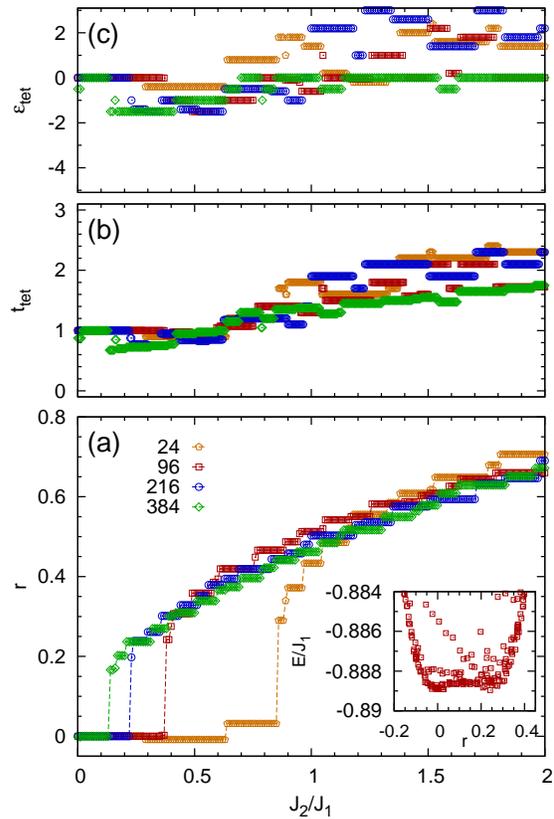}
 \caption{\label{fig:j2_r}
 (a) The tetramerization order parameter as a function of $J_2/J_1$ in the model with next nearest exchange $J_2$ for 24, 96, 216, and 384 site clusters.
A weakly first-order phase transition occurs between the algebraic color liquid and the tetramerized phase. The value of the $J_2/J_1$ decreases with increasing the system size.  
In the inset the energy vs. order parameter is shown for the calculated $t_{\text{tet}}$ and  $\varepsilon_{\text{tet}}$ values close to the transition point for the 96 site cluster. In (b) and (c) we show the values of $t_{\text{tet}}$ and $\varepsilon_{\text{tet}}$  giving the minimal energy solution in the parameter set we considered.
 }
\end{figure}

Extending the Hamiltonian (\ref{eq:HSU4}) with the next nearest neighbor exchange $J_2$, the number of bonds within a singlet tetramer increases [$c$ in Fig.~\ref{fig:ordpar_cube}(c)], but so does the number of bonds between the singlets [$d$ and $e$ in Fig.~\ref{fig:ordpar_cube}(c)] as well. 
The energy per site is now given by
\begin{equation}
 E = \frac{3}{2} J_1 \langle P_1 \rangle + 3 J_2 \langle P_2 \rangle \;,
 \label{eq:Ennn}
\end{equation}
where $J_1$ is the nearest neighbor exchange coupling, $\langle P_1 \rangle$ is defined in Eq.~(\ref{eq:P1def}), and 
\begin{equation}
  \langle P_2 \rangle = \frac{1}{4} \left( \langle P_c \rangle + 2 \langle P_d \rangle  + \langle P_e \rangle \right)
 \label{eq:P2def}
\end{equation}
is the average second neighbor exchange energy.

The answer to the question if the $J_2$ drives the system toward tetramerization is not evident. In the limiting case of the tetramer-factorized wave function, the expectation value of the energy is $E = -9 J_1/16 -3 J_2/16 = -0.5625 J_1 -0.1875 J_2$. This energy is to be compared with the expectation value of the energy in the algebraic color liquid, where $\langle P_a \rangle = \langle P_b \rangle \approx -0.596$ and $\langle P_c \rangle = \langle P_d \rangle = \langle P_e \rangle \approx 0.005$, so that $E \approx -0.895 J_1 + 0.015 J_2$. Comparing these two energies may serve as a rough estimate and we learn that  tetramerization becomes favorable for $J_2 \agt 1.6 J_1$. To get a better estimate of the transition, we searched for the minimum energy as we increased the $J_2$, using the tabulated values of $\langle P_1 \rangle$ and $\langle P_2 \rangle$ for different values of $t_{\text{tet}}$ and  $\varepsilon_{\text{tet}}$, and cluster sizes up to 384 sites. The results are shown in Fig.~\ref{fig:E_vs_r_J2} for $J_2/J_1=0.1$ and $0.2$, where we can clearly see the two local energy minima and the strong finite size dependence of the local minimum at finite $r\approx 0.2$. The order parameter as a function of $J_2/J_1$ is shown in Fig.~\ref{fig:j2_r}. The $J_2/J_1$ value of the first order transition between the tetramerized and algebraic liquid phase decreases with the system size, similarly to the case discussed in Fig.~\ref{fig:alpha_r}. From the plot it may appear that the $J_2=0$ is within the tetramerized phase in the thermodynamic limit. Since the precise value of the  transition depends on small energy differences, we believe that the finite size analysis we presented in Fig.~\ref{fig:ctuning} is more reliable.

Regarding the nature of the transitions, the two local minima in the energy-order parameter plot (the inset in Fig.~\ref{fig:j2_r}) are much less pronounced than in the case when the tetramerization was induced by the four-spin term $Q$ (see the corresponding inset in Fig.~\ref{fig:alpha_r}), and, as a consequence, the first order character of the phase transition is weaker.

In the case of very strong $J_2/J_1$ ({\it i.e.} $J_1 \rightarrow 0$), we get two decoupled triangular lattices. The SU(4) problem on the triangular lattice was studied in Ref.~\cite{penc2003}, where a plaquette state of resonating SU(4) singlets is proposed as the ground state, with a large (12 site) unit cell. Thus we may expect another phase transition as the $J_2/J_1$ is increased.   

\section{Conclusions}
\label{sec:Conclusions}

 In this paper we studied the tendency toward tetramerization -- SU(4) singlet plaquette formation -- in the ground states of the SU(4) Heisenberg models on honeycomb lattice. In contrast to the algebraic color (or spin-orbital) liquid discussed in Ref.~\cite{Corboz12_su4}, the tetramerized state breaks translational symmetry, resulting in a four-fold degenerate ground state. Tetramerization appears as a natural instability of the algebraic color liquid as described by the Gutzwiller projected $\pi$--flux state, since the wave vector associated with this distortion is compatible by the distance  between the Fermi points in Brillouin zone of the underlying free fermionic system.  
 
As a first step, we proved that the tetramerized state is a good ground state candidate: by adding four-site terms of the form $P_{ij}P_{kl}$ to the Heisenberg Hamiltonian, we get a model that is a sum of projection operators, and for which the product wave function of four-site SU(4) singlets is an exact ground state.
 
 Next, to detect and characterize the strength of tetramerization, we constructed a three dimensional order parameter based on symmetry group analysis. Furthermore,  we derived the corresponding symmetry allowed Landau-like free energy, where the   existence of the cubic term tells us that the phase transition between the tetramerized and the algebraic liquid state is of the first order. 
 
 Both the algebraic liquid and the tetramerized state can be described using a variational approach. In Ref.~\cite{Corboz12_su4} it was shown that the correlations in the algebraic liquid, as compared to exact diagonalization result from 24 site cluster, are nicely reproduced by the Gutzwiller projected free fermion wave function at quarter filling. The free fermions are hopping on the honeycomb lattice with amplitudes which comprise a $\pi$ flux in each hexagon (the product of the six hopping amplitudes around each hexagon is $-1$). Modulating the hopping amplitudes $t_{\text{tet}}$  between the tetramers and on--site energies $\varepsilon_{\text{tet}}$ in the center of tetramers, this free fermion wave function function can be continuously deformed to describe the tetramerized wave function. In particular, Gutzwiller projecting free fermions on decoupled tetramers is exactly the SU(4) singlet plaquette product wave function, the exact ground state for the Hamiltonian with the four-site term. 
 Sampling the Gutzwiller projected wave function by a Monte Carlo procedure, we calculated the expectation values of the exchange operators on the nonequivalent nearest and next nearest neighbor bonds, as well as of the four site term for a large number of different values of $t_{\text{tet}}$ and $\varepsilon_{\text{tet}}$.
 
   For the Heisenberg model with nearest neighbor exchange only, we observed that on the energy vs. order parameter plot most of the points closely follow a single curve even for quite different $t_{\text{tet}}$ and $\varepsilon_{\text{tet}}$ values, with some deviations for small values of $t_{\text{tet}}$. This indicates some kind of robustness of the variational wave function on the details of the hopping Hamiltonian. Furthermore, in the $(t_{\text{tet}} , \varepsilon_{\text{tet}})$ plane the energy is minimal along a narrow valley that closely follows the line of the formation of the Dirac point in the free fermion band structure, including also the point with vanishing tetramerization. All this emphasizes the significance of having Fermi-points instead of a gapped spectrum for the algebraic liquid. As for the finite size effects, we plotted the energy vs.\ order parameter for several cluster sizes: while the finite size effects are negligible for the non-tetramerized case, they become significant for the moderately tetramerized wave function. To confirm the stability of the algebraic spin liquid against tetramerization, we studied the finite size corrections of the coefficients in the free energy, and confirmed that the algebraic color liquid remains the ground state. 

 We also studied the tetramerization after introducing the four-site term or the next neighbor exchange term. Two local minima in the energy curve, associated with the first order phase transition, are clearly seen. For both added terms there were important finite size effects regarding the values of phase transition points, allowing for the possibility that the transition point is shifted so drastically that it includes also the model with the nearest neighbor exchange only. Clearly, this contradicts the analysis that we performed based on the finite size effect of the coefficients in the Landau free energy. Since in case of the first order phase transition the transition points are determined by tiny energy differences (that may depend on the details of calculations, like boundary conditions), we believe that we shall be cautious when performing finite size scaling of the transition point. 
 
 Independently of the precise value of the transition point,  our calculations show that the tetramerized SU(4) singlet plaquette state, next to the algebraic color liquid state, appears to be a strong ground state candidate for the four-component Mott insulating state on the honeycomb lattice at zero temperature, and its realization may depend on fine details of the effective Heisenberg model. 
  
 Certainly, it would be interesting to compare the results of the variational approach to those of  the mean field theory. In the case of the four site term, the mean field would introduce 3rd and 4th-neighbor hoppings, and the next-nearest exchange introduces next nearest neighbor hoppings. 
Introducing  longer range hoppings in the variational approach is straightforward, but the energy minimization would require more sophisticated methods where the hopping parameters are optimized by a stochastic minimization, as e.g. in Refs.~\cite{Yunoki2006,Iqbal2011} for the SU(2) Heisenberg model.

\acknowledgments

The authors thank P.~Corboz, A.~M.~L\"auchli, F.~Mila, and G.~Szirmai for stimulating discussions. We are grateful for the support of the Hungarian OTKA Grant No. 106047 and for the computational time on the Supercomputing Center of the ISSP, The University of Tokyo.
\appendix
\bibliographystyle{apsrev4-1}
\bibliography{refs}

\end{document}